\documentclass[reprint,prl,twocolumn,superscriptaddress,showpacs,showkeys,floatfix,preprintnumbers]{revtex4-1}

\usepackage{braket}
\usepackage{xargs}
\usepackage{mathtools}
\usepackage[english]{babel}
\usepackage[utf8]{inputenc}
\usepackage{amsmath,amssymb,braket,slashed,mathrsfs}
\usepackage{pifont}
\usepackage{MnSymbol}
\usepackage{graphicx}
\usepackage{tikz}
\usetikzlibrary{shapes,arrows,positioning}
\tikzset{decision/.style={diamond, draw, fill=blue!20, text width=4.5em, text badly centered, inner sep=0pt}}
\tikzset{block/.style={rectangle, draw, fill=blue!20, text width=10em, text centered, rounded corners, minimum width=3.5cm}}
\tikzset{block1/.style={rectangle, draw, fill=blue!20, text width=18.5em, text centered, rounded corners, minimum width=3.5cm}}
\tikzset{line/.style={draw, -latex, thick}}
\usepackage{color,hyperref}
\usepackage{multirow,array}
\usepackage{physics}

\usepackage{cleveref}
\crefformat{equation}{Eq.~(#2#1#3)}
\crefrangeformat{equation}{Eqs.~(#2#1#3)--(#2#4#3)}
\newcommand{\be}{\begin{equation}}
	\newcommand{\ee}{\end{equation}}
\newcommand{\ba}{\begin{eqnarray}}
\newcommand{\ea}{\end{eqnarray}}

\newcommand{\nn}{\nonumber}

\renewcommand{\vec}[1]{\mathbf{#1}}

\newcommand{\piz}{\pi^{0}}

\newcommand{\Gpigg}{\Gamma(\piz\!\to\!\gamma\gamma)}

\newcommand{\innovation}{Collaborative Innovation Center of Quantum Matter, Beijing 100871, China}
\newcommand{\chep}{Center for High Energy Physics, Peking University, Beijing 100871, China}
\newcommand{\pkuphy}{School of Physics, Peking University, Beijing 100871,
	China}

\newcommand{\Uconn}{Department of Physics, University of Connecticut, Storrs, CT 06269, USA}

\newcommand{\SCNT}{Southern Center for Nuclear-Science Theory (SCNT), Institute of Modern Physics, Chinese Academy of Sciences, Guangdong 516000, China}

\begin{document}
	\title{First-principles determination of anomaly-induced pion decay beyond the chiral limit}
	
	\author{Tian~Lin}\affiliation{\pkuphy}
	\author{Xu~Feng}\affiliation{\pkuphy}\affiliation{\innovation}\affiliation{\chep}\affiliation{\SCNT}
	\author{Lu-Chang Jin}\affiliation{\Uconn}
	\author{Chuan Liu}\affiliation{\pkuphy}\affiliation{\innovation}\affiliation{\chep}
	\author{Qi-Yuan Luo}\affiliation{\pkuphy}
	
	\date{\today}
	
	\begin{abstract}
	
		The two-photon decay of the neutral pion is fixed in the chiral limit by
	the Adler-Bell-Jackiw anomaly, 
while nonzero quark masses induce few-percent corrections that must be determined for precision tests of QCD beyond the chiral limit.
	Using the anomalous PCAC relation, we compute them in lattice QCD as deviations from the exact anomaly condition at $q^2=0$, thereby avoiding both
	four-point functions and the cancellation of chiral logarithms that limits
	conventional approaches. 
	Because the correction is proportional to the light-quark masses, both statistical and systematic uncertainties are correspondingly suppressed, making a precision calculation feasible.
	On two nearly-physical domain-wall ensembles we achieve $\sim1\%$ statistical precision each for the decay width, obtaining $\Gpigg=8.09(22)$~eV after continuum extrapolation.
	The $2.3(1.4)\%$ mass correction to the decay amplitude, 
	positive and isospin-breaking dominated, provides the first \textit{ab initio}
	confirmation of the $\piz$-$\eta$-$\eta'$ mixing enhancement.

	\end{abstract}
	
	\maketitle
	
\paragraph{Introduction} 

The neutral pion $\piz$ is the lightest hadron and plays a singular role in our understanding of quantum chromodynamics (QCD).
With a mean lifetime of only $\tau \approx 8.4 \times 10^{-17}$~s,
it decays almost exclusively ($\approx 98.8\%$ branching ratio) into
two photons via the electromagnetic interaction~\cite{ParticleDataGroup:2024cfk}.
Far from being a mundane low-energy process, this decay is
governed entirely by the chiral anomaly---a purely quantum
mechanical phenomenon in which a symmetry of the classical QCD
Lagrangian is violated by quantum fluctuations of the quark and gluon
fields~\cite{Adler:1969gk,Bell:1969ts}, 
causing $\pi^0$ to decay on a timescale about eight orders of magnitude shorter than that of $\pi^\pm$.
As a consequence, $\Gpigg$ serves as one of the most
fundamental and precisely testable predictions of quantum field
theory, 
making the precise measurement and theoretical computation of $\Gpigg$ a cornerstone test of QCD.

In the chiral limit with massless up and down quarks, the chiral anomaly yields an exact, parameter-free prediction for $\Gpigg$
\begin{equation}
  \Gamma(\piz\!\to\!\gamma\gamma)^\text{ABJ}
  = \frac{\pi m_{\pi}^3\alpha^2 K_0^2}{4 F_{\pi}^2}
  = 7.743(16)\ \mathrm{eV},
\label{eq:LO}
\end{equation}
where $K_0 = N_c(Q_u^2-Q_d^2)/(4\pi^2) = 1/(4\pi^2)$ is the ABJ anomaly
constant, with $Q_u=2/3$ and $Q_d=-1/3$. Here
$\alpha$ is the fine-structure constant, $m_{\pi}$ is the $\piz$
mass, $N_c = 3$ is the number of colors, and $F_{\pi} = 92.320(97)$~MeV
is the pion decay constant extracted from the weak decay of charged
pions~\cite{ParticleDataGroup:2024cfk}. 
This is essentially unique in QCD: almost all other low-energy
observables require either phenomenological model input or
nonperturbative numerical methods to compute.
Eq.~(\ref{eq:LO}) provides a direct link between the non-Abelian
gauge structure of QCD (through $N_c$), electroweak physics (through
$\alpha$ and $F_\pi$ from weak decays), and a strong-interaction
observable, making it a rare, exact bridge across sectors of the
Standard Model.

The real world contains quarks with finite masses, and these
introduce corrections to Eq.~(\ref{eq:LO}). Based on Chiral
Perturbation Theory (ChPT), the dominant effect arises from $\piz$-$\eta$-$\eta'$ mixing driven by
isospin breaking ($m_u \neq m_d$), which enhances the decay width
by approximately $4.5\%$ relative to the chiral limit~\cite{Goity:2002nn,Ananthanarayan:2002kj,Kampf:2009tk}.
On the experimental side, a dramatic improvement came from the PrimEx program at Jefferson Lab.
Combining PrimEx-I~\cite{PrimEx:2010fvg} and PrimEx-II~\cite{PrimEx-II:2020jwd} phases yields the most precisely measured
$\Gpigg$ to date:
\begin{equation}
  \Gpigg^\text{exp} = 7.802(117)\ \text{eV},
  \label{eq:primex}
\end{equation}
with a total uncertainty of 1.5\%. 
Crucially, this result is in agreement with the ABJ
anomaly prediction~(\ref{eq:LO}) but shows a 1.8-$\sigma$
tension with the NNLO ChPT prediction~\cite{Kampf:2009tk}.
An energy upgrade to 22 GeV of electron beam at JLab will enable measurements of $\Gpigg$ with sub-percent uncertainties~\cite{Accardi:2023chb},
making the quantitative understanding of $\Gpigg$ both timely and essential.

Lattice QCD is uniquely positioned to provide a systematically improvable, first-principles
computation of QCD observables without relying on model assumptions
or truncating a perturbative expansion. 
The first lattice QCD calculation of $\Gpigg$ was performed in 2012~\cite{Feng:2012ck}, shortly after PrimEx-I experiment released
its measurements. So far, various calculations have appeared with different methodologies~\cite{Gerardin:2016cqj,Gerardin:2019vio,Christ:2022rho,Gerardin:2023naa,ExtendedTwistedMass:2023hin,Koponen:2023zle,Lin:2024khg,Su:2026xdt}. 
These calculations determine the full decay amplitude directly in the isospin-symmetric (IS) limit, $m_u=m_d$, and therefore cannot access the isospin-breaking (IB) contributions 
that ChPT predicts to dominate the mass correction.
Moreover, their uncertainties, at the level of $4\%$-$12\%$, span a range covering both the ABJ anomaly prediction and the ChPT expectation, making the small mass correction difficult to resolve. 
The central values of some calculations even lie below the ABJ prediction, in contrast to the $\sim4.5\%$ enhancement expected from ChPT; see the Supplemental Material (SM)~\cite{SM}.

Rather than simply devoting more computational effort to the full decay amplitude, we change the quantity computed on the lattice 
and determine the mass corrections themselves, including both IS and IB contributions.
Since both are only at the few-percent level,
determining them to $\sim10\%$ already yields a sub-percent determination of $\Gpigg$.
A straightforward lattice implementation, however, requires introducing the IS and IB quark-mass operators perturbatively, converting the usual three-point functions into four-point functions. Besides the increased computational cost, this approach suffers from two major challenges. First, the IS operator $\bar{u}u+\bar{d}d$ generates disconnected diagrams with poor signal-to-noise ratios. Second, the mass derivatives of pion decay constant and the $\pi^0\to\gamma\gamma$ transition form factor,
$dF_\pi/dm_\pi^2$ and $dF_{\pi^0\gamma\gamma}/dm_\pi^2$,
each contain chiral logarithms that cancel only after the two contributions are combined. Computing them separately therefore leads to large logarithmically enhanced contributions that largely cancel in the final result, while the associated statistical uncertainties do not cancel correspondingly, resulting in substantially amplified relative uncertainties. To overcome these difficulties, we develop a novel method to compute
the IS and IB corrections directly through three-point
functions as described below, which 
reduces the uncertainties induced by the IS and IB contributions to the total $\Gpigg$ to 0.5-0.7\% and 0.8-1.0\%, respectively, on the used gauge ensembles.

\paragraph{Methodology} 
Introducing the isovector axial current
$A_\mu^3 = \frac{1}{2}(\bar u\gamma_\mu\gamma_5 u - \bar d\gamma_\mu\gamma_5 d)$
and the pseudoscalar bilinears
$P^3 = \frac{1}{2}(\bar u \gamma_5 u - \bar d \gamma_5 d)$ and
$P^0 = \frac{1}{2}(\bar u \gamma_5 u + \bar d \gamma_5 d)$ in Euclidean space,
the anomalous PCAC relation in QCD reads
\begin{equation}
  \partial_\mu A_\mu^3 = \hat{P}^3 + \hat{P}^0
  - \pi K_0 \alpha F^\mathrm{em}_{\mu\nu}\tilde{F}^{\mathrm{em}}_{\mu\nu},
  \label{eq:pcac}
\end{equation}
with $\hat{P}^3=2m_lP^3$ and $\hat{P}^0=\delta m P^0$,
where $m_l=(m_u{+}m_d)/2$, $\delta m=m_u{-}m_d$, and
$\tilde{F}_{\mu\nu}^{\mathrm{em}}=\frac{1}{2}\epsilon_{\mu\nu\rho\sigma}F_{\rho\sigma}^{\mathrm{em}}$. 
Throughout this paper, we use the notation $\hat{}$ for operators and correlators
carrying explicit quark-mass insertions, while ordinary operators and correlators are written without it.

The scalar invariants $\mathcal{W}$ and $\hat{\mathcal{F}}^{(a)}$ ($a=3,0$)
are defined by projecting the axial-vector-vector (AVV) amplitude
$\langle T[J^\mathrm{em}_\rho J^\mathrm{em}_\sigma A^3_\mu]\rangle$
and the pseudoscalar-vector-vector (PVV) amplitudes
$\langle T[J^\mathrm{em}_\rho J^\mathrm{em}_\sigma \hat{P}^a]\rangle$
onto the $\epsilon_{\rho\sigma\lambda\alpha}p_\lambda q_\alpha$ tensor structure
(with momenta $p$ carried by one photon and $q$ carried by $A_\mu^3$ or $\hat{P}^a$).
The Wick topologies of the PVV three-point functions are summarized in the SM~\cite{SM}.
Eq.~(\ref{eq:pcac}) then yields the Ward identity
\begin{equation}
  \mathcal{W}(-q^2) = \hat{\mathcal{F}}^{(3)}(-q^2) + \hat{\mathcal{F}}^{(0)}(-q^2) - K_0.
  \label{eq:ward}
\end{equation}
Details on the AVV and PVV amplitudes are provided in the SM~\cite{SM}.

With the momentum set
\be
q=(iE,\vec{0}),\quad p=\left(i\frac{E}{2},\vec{p}\right),\quad |\vec{p}|=\frac{E}{2},
\ee
$\hat{\mathcal{F}}^{(a)}(-q^2)$ can be evaluated as
\be
\label{eq:Fa}
\hat{\mathcal{F}}^{(a)}(-q^2)=\int d^4x\int d t_z\, \frac{e^{-Et_z}}{E}\frac{j_1\left(\frac{E}{2}|\vec{x}|\right)}{E|\vec{x}|}\hat{H}^{(a)}(t_z;x),
\ee
with the three-point correlator in Euclidean space
\begin{eqnarray}
  &&\hat{H}^{(a)}(t_z;x) = \epsilon_{\mu\nu\alpha 0}x_\alpha\times
\nn\\
&&\hspace{1cm}\int\! d^3\mathbf{z}\,
  \bigl\langle T\bigl[J^\mathrm{em}_\mu({\textstyle\frac{x}{2}})
  J^\mathrm{em}_\nu(-{\textstyle\frac{x}{2}})\hat{P}^{(a)}(\mathbf{z},t_z)\bigr]\bigr\rangle
  \label{eq:ht}
\end{eqnarray}
as inputs.
Since the AVV amplitude is regular at $q^2=0$ for nonzero pion mass,
$\mathcal{W}(0)=0$ gives the exact anomaly boundary condition
\begin{equation}
  \hat{\mathcal{F}}^{(3)}(0) + \hat{\mathcal{F}}^{(0)}(0) = K_0,
  \label{eq:bc}
\end{equation}
which holds to all orders in the strong coupling by anomaly non-renormalization~\cite{Adler:1969er}.

Rather than using the chiral-limit relation directly, we take Eq.~(\ref{eq:bc})
as the reference point. At the pion pole $-q^2=m_\pi^2$, the decay width is
\be
\label{eq:decay_width_pion_pole}
\Gamma(\pi^0\to\gamma\gamma)=\frac{\pi m_{\pi}^3\alpha^2 K^2}{4 F_{\pi}^2},
\ee
where
\ba
  \label{eq:kcalig}
K&=&\lim_{q^2+m_\pi^2\to 0}\frac{q^2+m_\pi^2}{m_\pi^2}\left( \hat{\mathcal{F}}^{(3)}(-q^2) + \hat{\mathcal{F}}^{(0)}(-q^2) \right)
\nn\\
&=&\hat{\mathcal{K}}^{(3)}(m_\pi^2)+\hat{\mathcal{K}}^{(0)}(m_\pi^2).
\ea
with $\hat{\mathcal{K}}^{(a)}(-q^2)\equiv\frac{q^2+m_\pi^2}{m_\pi^2}\hat{\mathcal{F}}^{(a)}(-q^2)$, which is regular at $-q^2=m_\pi^2$.

The mass correction is therefore
\ba
  \Delta K &=& K - K_0
\nn\\
  &=& \hat{\mathcal{K}}^{(3)}(m_\pi^2)+\hat{\mathcal{K}}^{(0)}(m_\pi^2) - \hat{\mathcal{K}}^{(3)}(0)-\hat{\mathcal{K}}^{(0)}(0).
  \label{eq:dkis}
\ea
In ChPT, electromagnetic corrections to the decay amplitude start at $\mathcal{O}(e^2p^4)$ and have been estimated to be about $0.3\%$~\cite{Ananthanarayan:2002kj}. 
In our formulation, however, the anomaly constraint~(\ref{eq:bc}) remains exact in the presence of QED. Consequently, electromagnetic effects cancel in the anomaly reference quantity, 
leaving only corrections of $\mathcal{O}(\alpha\, m_l)$ or $\mathcal{O}(\alpha\,\delta m)$ to $\Delta K$.
Thus, when both the light-quark mass difference and electromagnetic effects are regarded as IB effects, the residual electromagnetic contributions enter only at higher order in the IB expansion.

Separating the IS and IB contributions, the mass dependence of $\Delta K$ is decomposed as
\be
\Delta K=\Delta K^{\mathrm{IS}} +\Delta K^{\mathrm{IB}}
\ee
with
\ba
&&\Delta K^{\mathrm{IS}}=\left(\hat{\mathcal{K}}^{(3)}(m_\pi^2) -\hat{\mathcal{K}}^{(3)}(0)\right)\Big|_{\delta m=0}
\nn\\
&&\Delta K^{\mathrm{IB}}=-\delta m \left(\mathcal{K}^{(0)}(0)\Big|_{\delta m=0}\right)
\nn\\
&&\hspace{0.2cm}+\left(\hat{\mathcal{K}}^{(3)}(m_\pi^2) - \hat{\mathcal{K}}^{(3)}(0)\right)-\left(\hat{\mathcal{K}}^{(3)}(m_\pi^2) - \hat{\mathcal{K}}^{(3)}(0)\right)\Big|_{\delta m=0}
\nn\\
&&\hspace{0.2cm}+\delta m\left(\mathcal{K}^{(0)}(m_\pi^2)  - \left(\mathcal{K}^{(0)}(0)- \mathcal{K}^{(0)}(0)\Big|_{\delta m=0}\right)\right)
\nn\\
&&\hspace{0.95cm}=\Delta K^{\mathrm{IB},(0)}+\Delta K^{\mathrm{IB},(1)}+\Delta K^{\mathrm{IB},(2)}.
\ea
Here $\Delta K^{\mathrm{IB},(0)}$ is the leading IB contribution, while $\Delta K^{\mathrm{IB},(1)}$
and $\Delta K^{\mathrm{IB},(2)}$ arise at $\mathcal{O}(m_l\,\delta m)$ and $\mathcal{O}(\delta m^2)$, respectively.

The computation of $\hat{\mathcal{F}}^{(3)}(m_\pi^2)$ yields a pion-pole singularity.
This pion pole can be subtracted by defining
\begin{equation}
  \widetilde{H}^{(3)}(t_z,-t_s;x) = \hat{H}^{(3)}(t_z;x)
  - e^{m_\pi(t_s+t_z)}\hat{H}^{(3)}(-t_s;x),
  \label{eq:htilde}
\end{equation}
for a reference time $t_s$, which is chosen sufficiently large to guarantee the pion ground-state dominance in $\hat{H}^{(3)}(t_z;x)$ for $t_z\le -t_s$. 
Then the IS contribution can be given by
\ba
\label{eq:IS}
\Delta K^{\mathrm{IS}}&=&
\frac{2e^{m_\pi t_s}}{m_\pi^2} \int d^4x\left(\frac{j_1\left(\frac{m_\pi}{2}|\vec{x}|\right)}{m_\pi|\vec{x}|}-\frac{1}{6}\right)\hat{H}^{(3)}(-t_s;x)\Big|_{\delta m=0}
\nn\\
&&+\int d^4x\int_{-t_s}^0 dt_z\,\frac{t_z}{3} \widetilde{H}^{(3)}(t_z,-t_s;x)\Big|_{\delta m=0},
\ea
while the leading IB contribution is given by
\be
\label{eq:IB0}
\Delta K^{\mathrm{IB},(0)}=\delta m \int d^4x\int_{-\infty}^0 dt_z\,\frac{t_z}{3} \left(H^{(0)}(t_z,x)\Big|_{\delta m=0}\right).
\ee
The detailed derivations are given in the SM~\cite{SM}.

The subleading IB contributions $\Delta K^{\mathrm{IB},(1)}$ and $\Delta K^{\mathrm{IB},(2)}$ can be evaluated by explicitly inserting the IB mass operators. A complete calculation of these terms is substantially more complicated. Given that they are already suppressed relative to $\Delta K^{\mathrm{IB},(0)}$, we instead retain only the pion ground-state contribution to obtain an estimate of their magnitude. The detailed expressions used in this estimate are given in the SM~\cite{SM}.

\paragraph{Numerical results}

We use two $2+1$-flavor domain-wall fermion ensembles, 24D and 32Dfine, generated by the RBC-UKQCD Collaboration~\cite{RBC:2014ntl}. The two ensembles have nearly physical pion masses, $m_\pi=143.1(3)$ and $142.9(7)$ MeV~\cite{Lin:2024khg}, comparable spatial volumes ($L\simeq4.6$ fm), and different lattice spacings, $a^{-1}=1.023(2)$ and $1.378(5)$ GeV. 
We use 92 configurations for 24D and 49 for 32Dfine. Although we previously computed the $\pi^0\to\gamma\gamma$ amplitude using Coulomb-gauge wall-source pion operators~\cite{Lin:2024khg}, the PCAC formulation adopted here requires local pion interpolating operators. We therefore reconstruct the contractions using point-source propagators, with 1024 randomly chosen space-time origins per gauge configuration. 
The resulting $\pi^0\to\gamma\gamma$ amplitude agrees with the wall-source result within statistical uncertainties.
For the disconnected IB contributions, we additionally employ stochastic volume-source propagators, using 64 sources per configuration on 24D and 256 on 32Dfine. Combining the gauge configurations with the stochastic sources, we accumulate roughly $5,000$-$10,000$ measurements per ensemble for IB contributions.
Both electromagnetic currents are implemented as local vector currents, with renormalization factors taken from Ref.~\cite{Feng:2021zek}.
The evaluation of the $\pi$-state contribution to $\Delta K^{\mathrm{IB},(1)}$ requires four-point correlation functions. For these, the pion interpolating operator is constructed from a Coulomb-gauge-fixed wall source, while the isospin-breaking scalar density and one electromagnetic current are inserted using point-source propagators; the second current is treated as the sink operator.

The calculation requires both bare quark mass $m_l$ and $\delta m$. The lattice ensembles are generated in the isospin limit, where $m_l$ includes the residual mass contribution from domain-wall fermions. To match the convention of the isospin-symmetric pion mass, we rescale the quark-mass insertion by ${m_\pi^{\mathrm{iso}}}^2/m_\pi^2$,
 where $m_\pi^{\mathrm{iso}}=0.135$ GeV denotes the pion mass in the isospin-symmetric world~\cite{RBC:2014ntl,RBC:2018dos}, and $m_\pi$ is the lattice-determined pion mass. The $\delta m$ is determined from the PDG input $m_{u,d}^{\overline{\mathrm{MS}}}$ at the renormalization scale $\mu=2$ GeV and converted to the bare lattice parameter through
\be
\delta m=2m_l\frac{m_u^{\overline{\mathrm{MS}}}-m_d^{\overline{\mathrm{MS}}}}{m_u^{\overline{\mathrm{MS}}}+m_d^{\overline{\mathrm{MS}}}}.
\ee
The PDG input used is $m_u^{\overline{\mathrm{MS}}}/m_d^{\overline{\mathrm{MS}}}=0.462(12)$, corresponding to $(m_u^{\overline{\mathrm{MS}}}-m_d^{\overline{\mathrm{MS}}})/(m_u^{\overline{\mathrm{MS}}}+m_d^{\overline{\mathrm{MS}}})=-0.368(11)$~\cite{ParticleDataGroup:2024cfk}.
The FLAG average, $m_u^{\overline{\mathrm{MS}}}/m_d^{\overline{\mathrm{MS}}}=0.485(20)$~\cite{FlavourLatticeAveragingGroupFLAG:2024oxs}, differs from the PDG value by about $1\sigma$. This difference is included as a systematic uncertainty in our final results.

\begin{figure}[htbp]
\centering
\includegraphics[width=\columnwidth]{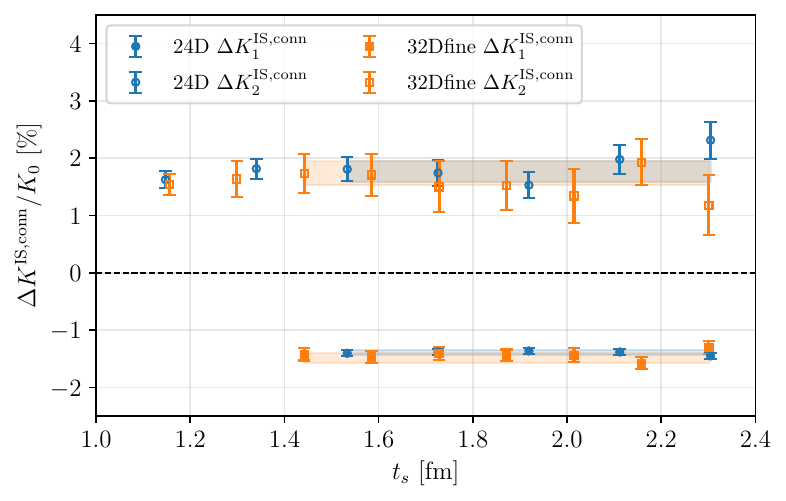}
\caption{The two terms of Eq.~(\ref{eq:IS}),
$\Delta K^{\mathrm{IS}}_1$ and
$\Delta K^{\mathrm{IS}}_2$ as a function of the
reference time $t_s$, for the 24D
and 32Dfine ensembles. The bands indicate the plateau averages used
in the final analysis. The two terms cancel to about $90\%$.}
\label{fig:dK_IS}
\end{figure}

\begin{figure}[htbp]
\centering
\includegraphics[width=\columnwidth]{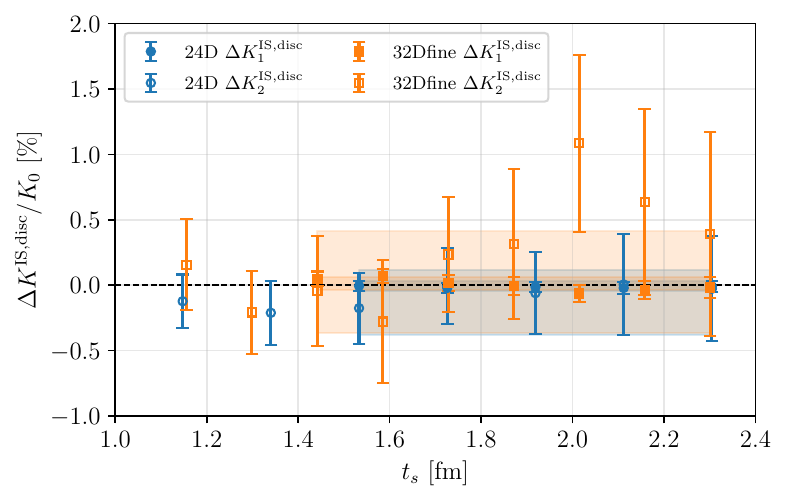}
\caption{Disconnected contribution $\Delta K^{\mathrm{IS,disc}}$, decomposed in
the same way as in Fig.~\ref{fig:dK_IS}. Both ensembles are consistent with
zero within the statistical precision.}
\label{fig:dK_IS_dis}
\end{figure}

We first discuss the IS contribution.
For connected contributions, $\Delta K^{\mathrm{IS},\mathrm{conn}}_1$ and $\Delta K^{\mathrm{IS},\mathrm{conn}}_2$ refer to the two terms in Eq.~(\ref{eq:IS}) and are shown in Fig.~\ref{fig:dK_IS} as a function of the
reference time $t_s$. The two terms are individually sizable and of opposite sign,
amounting to $-1.6\%$ and $+2.0\%$ of $K_0$, and each develops a clear
plateau for $t_s\gtrsim1.4$~fm on both ensembles, while their cancellation leads to a much smaller net contribution, namely
$\Delta K^{\mathrm{IS},\mathrm{conn}}/K_0=0.36(16)\%$ on 24D and
$0.31(19)\%$ on 32Dfine.
The disconnected contribution to the IS term, shown in
Fig.~\ref{fig:dK_IS_dis}, is the noisiest ingredient.
$\Delta K^{\mathrm{IS,disc}}/K_0=-0.14(20)\%$ (24D) and
$0.04(31)\%$ (32Dfine) are both consistent with zero.

\begin{figure}[htbp]
\centering
\includegraphics[width=\columnwidth]{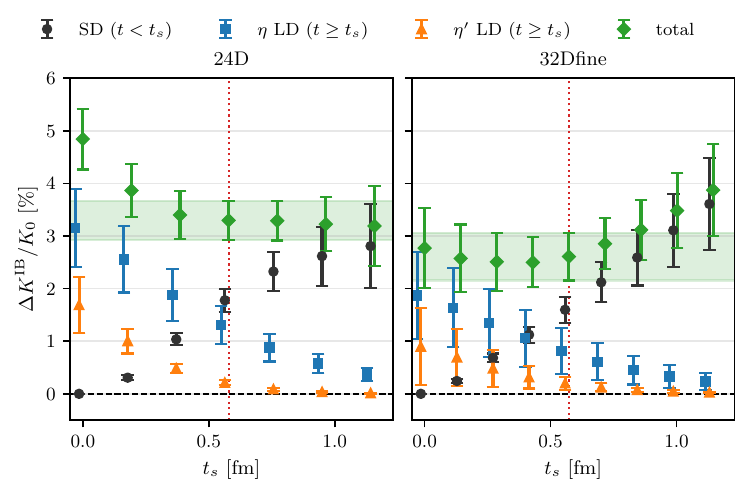}
\caption{Leading IB contribution $\Delta K^{\mathrm{IB}}$ as a function of the
matching time $t_s$ for 24D (left) and 32Dfine (right). Shown are the
short-distance lattice part ($t<t_s$), the $\eta$- and $\eta'$-dominated
reconstructed long-distance parts ($t\ge t_s$), and their total. The vertical
lines mark the chosen $t_s$; the horizontal bands give the quoted results.}
\label{fig:dK_IB}
\end{figure}

We now turn to the IB contribution, which dominates $\Delta K$. The leading
term $\Delta K^{\mathrm{IB},(0)}$ of Eq.~(\ref{eq:IB0}) involves the isoscalar
correlator $H^{(0)}$ integrated over the whole negative-time region. 
The integrand at large $|t_z|$ is relatively noisy. 
Instead of performing a temporal truncation,
we split
the integral at a matching time $t_s$: the short-distance region is taken directly
from the lattice correlator (SD), while the long-distance tail is reconstructed
from $\eta$ and $\eta'$ contributions. To this end, we perform a
combined two-state fit to the two-point correlation functions using wall-source operators. The interpolating basis consists of the light axial-vector current and the strange pseudoscalar density.
The two-point fit provides the $\eta^{(\prime)}$ masses and overlaps; with these parameters fixed on each jackknife sample, the matrix elements $\langle\eta^{(\prime)}|P^0|0\rangle$
and $\langle 0|T[J^\mathrm{em}J^\mathrm{em}]|\eta^{(\prime)}\rangle$ are obtained from uncorrelated linear fits. 
More details are given in the SM~\cite{SM}.
Fig.~\ref{fig:dK_IB} shows this decomposition. As $t_s$ increases the SD part
rises and the reconstructed long-distance (LD) parts fall, with the $\eta'$ LD
piece dying out fastest. We
quote the total at $t_s\simeq0.58$~fm (24D) and $0.57$~fm (32Dfine), marked by
the vertical lines. At this matching point the $\eta$ tail accounts for $40(8)\%$
(24D) and $31(13)\%$ (32Dfine) of the total, and the $\eta'$ tail for a further
$6(1)\%$ and $8(5)\%$, respectively. This gives
$\Delta K^{\mathrm{IB},(0)}/K_0=3.30(42)\%$ on 24D and
$2.61(48)\%$ on 32Dfine, roughly an order of
magnitude larger than the IS term.

The subleading IB terms are evaluated on 24D in the $\pi$-state
approximation described above. Both are found to be negligible,
$\Delta K^{\mathrm{IB},(1)}/K_0=-0.010(84)\%$ at
$\mathcal{O}(m_l\,\delta m)$ and
$\Delta K^{\mathrm{IB},(2)}/K_0=-0.0039(15)\%$ at
$\mathcal{O}(\delta m^2)$, i.e.\ two orders of magnitude below
$\Delta K^{\mathrm{IB},(0)}$ and statistically indistinguishable from zero. This
confirms the expected suppression pattern. We do not include their central
values, and instead retain their uncertainties as additional systematics on
the 24D $\Delta K$ total.

The individual contributions and their totals are collected in
Table~\ref{tab:dK}. Summing the IS and IB pieces together with the
disconnected pieces gives
$\Delta K/K_0=3.52(50)\%$ on 24D and $2.96(60)\%$ on 32Dfine.
The two ensembles differ by lattice spacing at nearly identical pion mass and
volume, and agree well within errors, with no obvious evidence for significant discretization effects at the current statistical precision. 
A linear extrapolation in $a^2$
yields the continuum value
\be
\frac{\Delta K}{K_0} = 2.3(1.4)\%,
\ee
whose enlarged uncertainty simply reflects the lever arm between the two
available lattice spacings.

\begin{table}[htbp]
\centering
\begin{tabular}{lccc}
\hline\hline
 & 24D & 32Dfine & continuum \\
\hline
$\Delta K^{\mathrm{IS,conn}}/K_0$~[\%] & $0.36(16)$ & $0.31(19)$ & $0.25(46)$ \\
$\Delta K^{\mathrm{IB}}/K_0$~[\%] & $3.30(42)$ & $2.61(48)$ & $1.8(1.1)$ \\
$\Delta K^{\mathrm{IS,disc}}/K_0$~[\%] & $-0.14(20)$ & $0.04(31)$ & $0.25(73)$ \\
$\Delta K^{\mathrm{IB,disc}}/K_0$~[\%] & $0.016(27)$ & $-0.003(42)$ & -- \\
$\Delta K^{\mathcal{O}(m_l\delta m)}/K_0$~[\%] & $-0.010(84)$ & -- & -- \\
$\Delta K^{\mathcal{O}(\delta m^2)}/K_0$~[\%] & $-0.0039(15)$ & -- & -- \\
\hline
$\Delta K/K_0$~[\%] & $3.52(50)$ & $2.96(60)$ & $2.3(1.4)$ \\
\hline
$\Delta\Gamma^{\mathrm{IS}}$~[eV] & $0.056(24)$ & $0.048(29)$ & $0.039(70)$ \\
$\Delta\Gamma^{\mathrm{IB}}$~[eV] & $0.511(65)$ & $0.404(74)$ & $0.27(17)$ \\
$\Delta\Gamma^{\mathrm{IS,disc}}$~[eV] & $-0.021(31)$ & $0.006(48)$ & $0.04(11)$ \\
$\Delta\Gamma^{\mathrm{IB,disc}}$~[eV] & $0.0025(42)$ & $-0.0005(65)$ & -- \\
\hline
$\Delta\Gamma$~[eV] & $0.545(77)$ & $0.458(93)$ & $0.35(22)$ \\
\hline\hline
\end{tabular}
\caption{Contributions to the quark-mass correction, quoted as the relative
shift $\Delta K/K_0$ of the anomaly constant, and to the corresponding
	decay-width shift $\Delta\Gamma$. The IS and IB terms use the
	$m_\pi^2$-rescaled central values and are combined before the errors are
	propagated; the IS-disconnected term enters with its central value, while the IB-disconnected contribution is an order of magnitude smaller and can be neglected; the two
	higher-order IB terms, computed on 24D only, enter through their
	uncertainties alone. The uncertainty in the input $m_u^{\overline{\mathrm{MS}}}/m_d^{\overline{\mathrm{MS}}}$ propagates to $K/K_0$, 
	giving an uncertainty of $0.1$-$0.2\%$, which is included in quadrature in the IB uncertainty.}
\label{tab:dK}
\end{table}

Since $\Delta K/K_0\ll1$, the width shift follows from
$\Delta\Gamma/\Gamma^\mathrm{ABJ}=2\Delta K/K_0$, giving
$\Delta\Gamma=0.35(22)$~eV in the continuum, i.e.\ a
$4.5(2.8)\%$ enhancement over the chiral-limit prediction. Adding this to
Eq.~(\ref{eq:LO}) yields our final result
\be
\Gamma(\piz\to\gamma\gamma)=8.09(22)\ \mathrm{eV},
\ee
where the error is dominated by the continuum extrapolation; the individual
ensembles give the more precise values $8.29(8)$~eV (24D) and $8.20(9)$~eV
(32Dfine).

\paragraph{Conclusion}
We have presented a first-principles lattice QCD determination of the
quark-mass corrections to the anomaly-induced decay $\piz\to\gamma\gamma$.
The central methodological point is that the anomalous PCAC relation allows the
correction to be defined relative to the exact anomaly condition
$\hat{\mathcal{F}}^{(3)}(0)+\hat{\mathcal{F}}^{(0)}(0)=K_0$ at $q^2=0$, which holds to all
orders in the strong coupling and also in the presence of QED. This
reformulation replaces the four-point functions required by a direct
mass-insertion calculation with three-point functions, and simultaneously
avoids the cancellation between $dF_\pi/dm_\pi^2$ and
$dF_{\pi^0\gamma\gamma}/dm_\pi^2$ that inflates the relative uncertainty 
when the two contributions are computed separately.

\begin{figure}[htbp]
\centering
\includegraphics[width=\columnwidth]{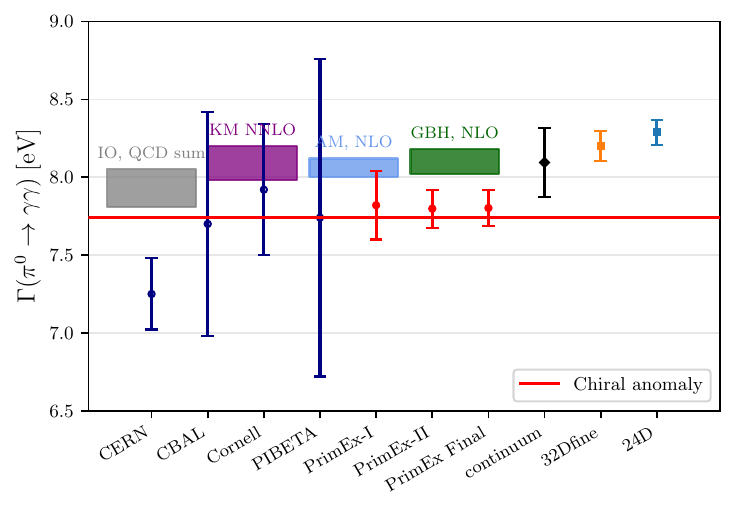}
\caption{Comparison of $\Gpigg$ from this work (24D, 32Dfine and continuum) with
experimental measurements~\cite{Atherton:1985av,CrystalBall:1988xvy,Browman:1974cu,Bychkov:2008ws,PrimEx:2010fvg,PrimEx-II:2020jwd} and with ChPT predictions (IO QCD sum rule~\cite{Ioffe:2007eg}; KM
NNLO~\cite{Kampf:2009tk}; AM NLO~\cite{Ananthanarayan:2002kj}; GBH
NLO~\cite{Goity:2002nn}). The horizontal line is the chiral-anomaly prediction of
Eq.~(\ref{eq:LO}).}
\label{fig:width}
\end{figure}

Fig.~\ref{fig:width} compares our lattice determination with the experimental results and ChPT predictions. We achieve a statistical precision of about $1\%$ for the $\pi^0\to\gamma\gamma$ decay width on each ensemble. The continuum extrapolation from two lattice spacings enlarges the final uncertainty to $2.7\%$. The correction is positive and dominated by isospin breaking, providing, for the first time, a lattice-QCD confirmation of the enhancement over the ABJ prediction anticipated from $\pi^0$-$\eta$-$\eta'$ mixing. Both the sign and magnitude agree with the ChPT predictions~\cite{Kampf:2009tk,Ananthanarayan:2002kj,Goity:2002nn}. The continuum result lies $1.2\sigma$ above the PrimEx I+II measurement~\cite{PrimEx-II:2020jwd}. 
A third lattice spacing is needed not only to strengthen the continuum extrapolation, but also to reduce the uncertainty in the comparison with experiment.

Since the correction we compute is proportional to the light-quark masses, $m_l$ or $\delta m$, its statistical and systematic uncertainties are correspondingly suppressed.
This is a key advantage over the conventional approach and is what makes the sign and magnitude of the small correction accessible with useful precision.
A dimensional estimate based on $a^2\Lambda_{\rm QCD}^2$ suggests relative discretization effects at the $\sim20\%$ level, consistent with the observed 
$\sim20\%$ difference between the 24D and 32Dfine results for $\Delta K/K_0$.
Eliminating this positive correction would require an $\mathcal O(100\%)$ discretization effect acting coherently at both lattice spacings.
Residual discretization effects are therefore unlikely to alter our conclusion of a positive, predominantly isospin-breaking mass correction.
Other systematic effects, including finite-volume corrections and the distinction between $F_{\pi^0}$ and $F_{\pi^\pm}$, are subleading at the current precision; see the SM~\cite{SM}.

Our methodology provides a direct path toward a future sub-percent determination of the decay width. Such a precision calculation can be confronted with the new measurements enabled by the 22 GeV upgrade at JLab~\cite{Accardi:2023chb}, opening the way to a precision test of the chiral anomaly beyond the chiral limit.

\begin{acknowledgments}
X.F., T.L., C.L. and Q.Y.L. were supported in
part by NSFC of China under Grants No. 12125501,
No. 12550007, No. 12293060 and No. 12293063.
L.J. acknowledges the support of DOE Office of Science
Early Career Award DE-SC0021147, DOE grant DE-
SC0010339 and DE-SC0026314. The research reported in
this work was carried out using the computing facilities
at Chinese National Supercomputer Center in Tianjin. It
also made use of computing and long-term storage facil-
ities of the USQCD Collaboration, which are funded by
the Office of Science of the U.S. Department of Energy.
\end{acknowledgments}

\bibliography{ref.bib}

\clearpage

\setcounter{page}{1}
\renewcommand{\thepage}{Supplementary Information -- S\arabic{page}}
\setcounter{table}{0}
\renewcommand{\thetable}{S\,\Roman{table}}
\setcounter{equation}{0}
\renewcommand{\theequation}{S\,\arabic{equation}}
\setcounter{figure}{0}
\renewcommand{\thefigure}{S\,\arabic{figure}}

\section{Supplementary Material}

\subsection{Lattice-QCD determinations of $\Gpigg$ in the isospin-symmetric limit}

\begin{figure}[htbp]
\centering
\includegraphics[width=\columnwidth]{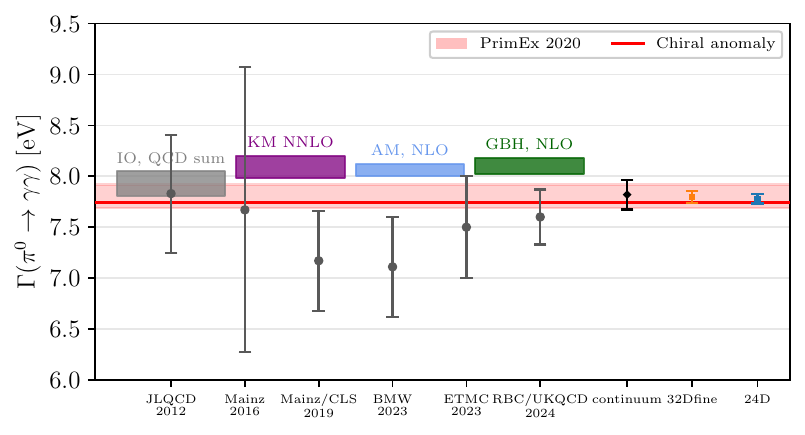}
\caption{Lattice QCD determinations of $\Gpigg$ in the IS limit, together with ChPT predictions (IO QCD sum rule~\cite{Ioffe:2007eg}; KM NNLO~\cite{Kampf:2009tk}; AM NLO~\cite{Ananthanarayan:2002kj}; GBH NLO~\cite{Goity:2002nn}) and the PrimEx I+II experimental measurement~\cite{PrimEx:2010fvg,PrimEx-II:2020jwd}. The lattice results include JLQCD (2012)~\cite{Feng:2012ck}, Mainz (2016)~\cite{Gerardin:2016cqj}, Mainz/CLS (2019)~\cite{Gerardin:2019vio}, BMW (2023)~\cite{Gerardin:2023naa}, ETMC (2023)~\cite{ExtendedTwistedMass:2023hin}, RBC/UKQCD (2024)~\cite{Lin:2024khg} and this work. For Mainz (2016) and Mainz/CLS (2019), $\Gpigg$ is obtained from the reported $\pi^0\to\gamma\gamma$ transition form factors; the other lattice results are taken directly from the respective references. The red horizontal line denotes the chiral-anomaly prediction of Eq.~(\ref{eq:LO}), and the pink band shows the combined PrimEx I+II measurement.}
\label{fig:lattice_width}
\end{figure}

Fig.~\ref{fig:lattice_width} first collects previous lattice QCD determinations (black dotted points) together with the ChPT predictions and the PrimEx I+II measurement. Since the lattice calculations were performed in the IS limit, their direct comparison with the physical result is incomplete, as they do not include IB effects. Nevertheless, the figure illustrates both the precision attainable in direct lattice calculations of the full decay amplitude and the spread of the resulting central values. It highlights the challenge of obtaining a lattice determination that can resolve the few-percent mass correction and directly discriminate between the ChPT prediction and experiment. This motivates our new strategy of computing the finite-mass correction itself, using the anomalous PCAC relation to reduce the required four-point functions to three-point functions and thereby achieve substantially improved precision.

For comparison, our IS result in the continuum limit is
\be
\Gamma(\pi^0\to\gamma\gamma)^{\mathrm{IS}}=7.82(13)\,\mathrm{eV}
\ee
with $7.78(4)$ and $7.80(6)$~eV on the 24D and 32Dfine ensembles, respectively. The corresponding transition form factor is
\be
F_{\pi^0\gamma\gamma}^{\mathrm{IS}}=0.2757(24)\,\mathrm{GeV}^{-1}.
\ee
where the experimental value $F_\pi=92.320(97)$~MeV is used as input. This is consistent with the FLAG consensus value $F_\pi^{\mathrm{IS}}=130.5/\sqrt{2}=92.277$~MeV~\cite{FlavourLatticeAveragingGroupFLAG:2024oxs}, differing by only $5\times10^{-4}$.

\subsection{Contraction topologies}
\label{sec:contractions}

\begin{figure}[htbp]
\centering
\includegraphics[width=\columnwidth]{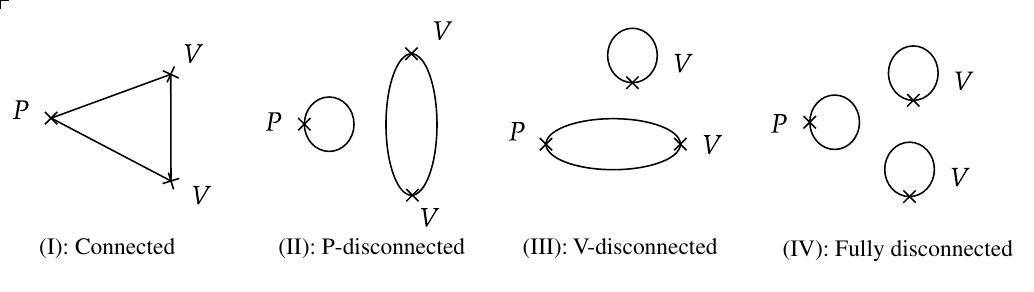}
\caption{
The four Wick topologies of the PVV three-point function:
(I) Connected, (II) P-disconnected, (III) V-disconnected, and (IV) Fully disconnected.
}
\label{fig:contractions}
\end{figure}

The four topologies are shown in Fig.~\ref{fig:contractions}.
For the isovector correlator $\langle T[J_\rho^{\mathrm{em}}J_\sigma^{\mathrm{em}}P^3]\rangle$, which enters the IS contribution, topologies (I) and (III) are included in our calculation. 
Topologies (II) and (IV) do not contribute in the isospin limit, since the pseudoscalar density $P^3$ forms an isolated quark loop in both cases. 
The corresponding $u$- and $d$-quark contributions are identical in the isospin limit and cancel in the $u-d$ combination defining $P^3$.
For the isoscalar correlator $\langle T[J_\rho^{\mathrm{em}}J_\sigma^{\mathrm{em}}P^0]\rangle$, which enters the IB contribution, topologies (I), (II) and (III) are included in our calculation. 
Topology (IV) is omitted because it is doubly suppressed by the SU(3)-flavor structure and, in addition, involves three disconnected quark loops, making it particularly noisy. Topology (III) in the IB contribution is suppressed relative to its IS counterpart by the charge factor $(Q_u+Q_d)/(Q_u-Q_d)=1/3$ and by the additional factor $\delta m/m_l$. It is therefore an order of magnitude smaller and
numerically negligible at our current precision

\subsection{Derivations of Eqs.~(\ref{eq:Fa}), (\ref{eq:IS}) and (\ref{eq:IB0})}

To turn the local identity~(\ref{eq:pcac}) into a relation among the amplitudes entering
\(\pi^0\to\gamma\gamma\), we first define the AVV and
PVV three-point functions. 
The axial and pseudoscalar amplitudes are the
direct Euclidean Fourier transform
\ba
  A^{(3)}_{\mu5\rho\sigma}(p,q)
  &=&
  -i\int d^4x\,d^4z\, e^{-ip\cdot x}e^{iq\cdot z}
\nn\\
  &&\left\langle
  T\left[
J^\mathrm{em}_\rho(x)J^\mathrm{em}_\sigma(0)A_\mu^3(z)
  \right]\right\rangle,
\nn\\
  A^{(a)}_{5\rho\sigma}(p,q)
  &=&
  \int d^4x\,d^4z\, e^{-ip\cdot x}e^{iq\cdot z}
\nn\\
&&
  \left\langle
  T\left[
J^\mathrm{em}_\rho(x)J^\mathrm{em}_\sigma(0)\hat{P}^a(z)
  \right]\right\rangle.
  \label{eq:pvv-correlator}
\ea
Parity and electromagnetic gauge invariance
\ba
p_\rho A^{(3)}_{\mu5\rho\sigma}(p,q) &=& 0
\nn\\
(p_\sigma - q_\sigma) A^{(3)}_{\mu5\rho\sigma}(p,q) &=& 0
\nn\\
p_\rho A^{(a)}_{5\rho\sigma}(p,q) &=& 0
\nn\\
(p_\sigma - q_\sigma) A^{(a)}_{5\rho\sigma}(p,q) &=& 0
\ea
strongly constrain the tensor structures of these amplitudes.
With these amplitude definitions, the scalar invariant functions are defined by
\ba
  &&A^{(3)}_{\mu5\rho\sigma}(p,q)
  =
  q_\mu \epsilon_{\rho\sigma\lambda\alpha}p_\lambda q_\alpha\,
  \mathcal{W}_1(-q^2)
\nn\\ &&\hspace{1cm}
+ \big[
  (\epsilon_{\mu\rho\lambda\alpha} p_\sigma - \epsilon_{\mu\sigma\lambda\alpha} (q_\rho - p_\rho))q_\lambda p_\alpha
\nn\\ &&\hspace{1.5cm}
  -
  \epsilon_{\mu\rho\sigma\lambda} (2 p_\lambda - q_\lambda) (q - p)\cdot p
  \big]
  \mathcal{W}_2(-q^2),
  \label{eq:w-projection-full}
\nn\\
  &&q^\mu A^{(3)}_{\mu5\rho\sigma}(p,q)
  =
  \epsilon_{\rho\sigma\lambda\alpha}p_\lambda q_\alpha\,
  \mathcal{W}(-q^2),
  \label{eq:w-projection}
\nn\\
  &&A^{(a)}_{5\rho\sigma}(p,q)
  =
  \epsilon_{\rho\sigma\lambda\alpha}p_\lambda q_\alpha\,
  \hat{\mathcal{F}}^{(a)}(-q^2).
  \label{eq:f-projection}
\ea
For the two-real-photon kinematics, only one invariant variable remains, namely $q^2$, since the two photon virtualities are fixed to zero. In this kinematics we have
\ba
  && \mathcal{W}(-q^2) =
  q^2 \mathcal{W}_1(-q^2) + q^2 \mathcal{W}_2(-q^2).
  \label{eq:w-relation}
\ea
For non-zero pion mass, there is no singularity in $\mathcal{W}_1(-q^2)$ and $\mathcal{W}_2(-q^2)$. This relation proves
\ba
\mathcal{W}(0) = 0.
\label{eq:w-condition}
\ea
This leads to the core relation \cref{eq:bc} of this work.
The anomaly amplitude arising from the
PCAC substitution takes the form
\[
  A^{\mathrm{anom}}_{\rho\sigma}(p,q)
  =
  -\epsilon_{\rho\sigma\lambda\alpha}p_\lambda q_\alpha\,K_0 .
\]

Defining $k=p-\frac{q}{2}=(0,\vec{p})$,  the scalar amplitude $\hat{\mathcal{F}}^{(a)}(-q^2)$ can be computed by projecting the PVV amplitudes
onto the $\epsilon_{\rho\sigma\lambda\alpha}k_\lambda q_\alpha$ tensor structure
\ba
\hat{\mathcal{F}}^{(a)}(-q^2)&=&\frac{\epsilon_{\rho\sigma\lambda\alpha}k_\lambda q_\alpha}{2(k^2q^2-(k\cdot q)^2)} 
\nn\\
&&\int d^4x\, e^{-ik\cdot x} \int dt_z\, e^{-Et_z} \,\hat{H}_{\rho\sigma}^{(a)}(t_z;x)
\nn\\
&=&2i\frac{\epsilon_{\rho\sigma \lambda 0}}{E^3} \int d^4x\, \left(i\frac{\partial }{\partial x_\lambda} j_0\left(\frac{E}{2}|\vec{x}|\right)\right)
\nn\\
&&\hspace{1cm}\int dt_z \, e^{-Et_z} \,\hat{H}_{\rho\sigma}^{(a)}(t_z;x)
\nn\\
&=& \int d^4x\int dt_z\,\frac{e^{-Et_z}}{E}\frac{j_1\left(\frac{E}{2}|\vec{x}|\right)}{E|\vec{x}|}\,\hat{H}^{(a)}(t_z;x)
\nn\\
\ea
where the hadronic tensor $\hat{H}_{\rho\sigma}^{(a)}(t_z;x)$ is defined as
\be
\hat{H}_{\rho\sigma}^{(a)}(t_z;x)=\int\! d^3\mathbf{z}\,
  \bigl\langle T\bigl[J^\mathrm{em}_\rho({\textstyle\frac{x}{2}})
  J^\mathrm{em}_\sigma(-{\textstyle\frac{x}{2}})\hat{P}^{(a)}(\mathbf{z},t_z)\bigr]\bigr\rangle.
\ee

Using the antisymmetry of $\hat{H}^{(a)}(t_z;x)$ under $t_z\to-t_z$, one finds
\ba
&&\hat{\mathcal{F}}^{(a)}(-q^2)
\nn\\
&=&-\int d^4x\int dt_z\,\frac{\sinh (Et_z)}{E} \frac{j_1\left(\frac{E}{2}|\vec{x}|\right)}{E|\vec{x}|}\,\hat{H}^{(a)}(t_z;x)
\nn\\
&=&-2\int d^4x\int_{-\infty}^0 dt_z\,\frac{\sinh (Et_z)}{E} \frac{j_1\left(\frac{E}{2}|\vec{x}|\right)}{E|\vec{x}|}\,\hat{H}^{(a)}(t_z;x).
\nn\\
\ea
For $\hat{\mathcal{F}}^{(0)}(-q^2)$, taking the limit of $E\to0$ yields Eq.~(\ref{eq:IB0}). 

For the computation of $\hat{\mathcal{F}}^{(3)}(-q^2)$,
the hadronic function $\hat{H}^{(3)}(t_z;x)$ is split into two terms
\be
\hat{H}^{(3)}(t_z;x)=  \widetilde{H}^{(3)}(t_z,-t_s;x) 
  + e^{m_\pi(t_s+t_z)}\hat{H}^{(3)}(-t_s;x),
\ee
and $\hat{\mathcal{F}}^{(3)}(-q^2)$ can be written as
\ba
&&\hat{\mathcal{F}}^{(3)}(-q^2)
\nn\\
&=&-2\int d^4x\int_{-\infty}^0 dt_z\,\frac{\sinh (Et_z)}{E} \frac{j_1\left(\frac{E}{2}|\vec{x}|\right)}{E|\vec{x}|}\,\widetilde{H}^{(3)}(t_z,-t_s;x) 
\nn\\
&&-2\int d^4x\,\frac{e^{m_\pi t_s}}{E^2-m_\pi^2} \frac{j_1\left(\frac{E}{2}|\vec{x}|\right)}{E|\vec{x}|}\hat{H}^{(3)}(-t_s;x)
\ea
where the Euclidean integral $\int_{-\infty}^0 dt_z\,\frac{\sinh(Et_z)}{E}\,e^{m_\pi t_z}$ in the second line is formally divergent. This divergence arises from the Euclidean correlator at large $|t_z|$ and is removed properly, yielding the factor $\frac{1}{E^2-m_\pi^2}$.
For fixed $t_x$ and sufficiently large $t_s>0$, ground-state dominance in $\hat{H}^{(3)}(t_z;x)$ for $t_z<-t_s$ is assumed. Under this condition, $\widetilde{H}^{(3)}(t_z,-t_s;x) =0$ and
\ba
&&\hat{\mathcal{F}}^{(3)}(-q^2)
\nn\\
&=&-2\int d^4x\int_{-t_s}^0 dt_z\,\frac{\sinh (Et_z)}{E} \frac{j_1\left(\frac{E}{2}|\vec{x}|\right)}{E|\vec{x}|}\,\widetilde{H}^{(3)}(t_z,-t_s;x) 
\nn\\
&&-2\int d^4x\,\frac{e^{m_\pi t_s}}{E^2-m_\pi^2} \frac{j_1\left(\frac{E}{2}|\vec{x}|\right)}{E|\vec{x}|}\hat{H}^{(3)}(-t_s;x).
\ea
At $E=m_\pi$ and $E=0$, we have
\ba
\hat{\mathcal{K}}^{(3)}(m_\pi^2)&=&\lim_{E\to m_\pi}\frac{m_\pi^2-E^2}{m_\pi^2}\hat{\mathcal{F}}^{(3)}(-q^2)
\nn\\
&=&\frac{2e^{m_\pi t_s}}{m_\pi^2} \int d^4x\,\frac{j_1\left(\frac{m_\pi}{2}|\vec{x}|\right)}{m_\pi|\vec{x}|}\hat{H}^{(3)}(-t_s;x)
\nn\\
\ea
and
\ba
\hat{\mathcal{K}}^{(3)}(0)
&=&-\int d^4x\int_{-t_s}^0 dt_z\,\frac{t_z}{3} \widetilde{H}^{(3)}(t_z,-t_s;x)
\nn\\
&&+\frac{2e^{m_\pi t_s}}{m_\pi^2} \int d^4x\,\frac{1}{6}\hat{H}^{(3)}(-t_s;x).
\ea
Combining the above two expressions yields Eq.~(\ref{eq:IS}).

\subsection{Subleading IB contributions}
To estimate the subleading IB contributions, we introduce the IB scalar operator
\be
S^3=\frac{1}{2}(\bar{u}u-\bar{d}d)
\ee
The $\pi$-state contribution to the $\mathcal{O}(m_l\,\delta m)$ IB term is 
\ba
\Delta K^{\mathrm{IB},(1)}&\approx&
\delta m\frac{2}{m_\pi^2} \int d^4x\int d^4y\left(\frac{j_1\left(\frac{m_\pi}{2}|\vec{x}|\right)}{m_\pi|\vec{x}|}-\frac{1}{6}\right)
\nn\\
&&\hspace{0.5cm}\times \frac{\epsilon_{\mu\nu\alpha 0}x_\alpha}{2m_\pi}
  \bigl\langle 0\bigl| T\bigl[J^\mathrm{em}_\mu({\textstyle\frac{x}{2}})
  J^\mathrm{em}_\nu(-{\textstyle\frac{x}{2}})S^3(y)\bigr|\pi\rangle
  \nn\\
  &&\hspace{0.5cm}\times \langle \pi|\hat{P}^{(3)}(0)|0\rangle.
\ea
Evaluating the matrix element
\be
\bigl\langle 0\bigl| T\bigl[J^\mathrm{em}_\mu({\textstyle\frac{x}{2}})
  J^\mathrm{em}_\nu(-{\textstyle\frac{x}{2}})S^3(y)\bigr|\pi\rangle
 \ee
requires the construction of a four-point correlation function.

The $\pi$-state contribution to the $\mathcal{O}(\delta m^2)$ IB term is
\ba
\Delta K^{\mathrm{IB},(2)}&\approx&
\delta m^2\frac{2}{m_\pi^2} \int d^4x\left(\frac{j_1\left(\frac{m_\pi}{2}|\vec{x}|\right)}{m_\pi|\vec{x}|}-\frac{1}{6}\right)
\nn\\
&&\hspace{0.5cm}\times \frac{\epsilon_{\mu\nu\alpha 0}x_\alpha}{2m_\pi}
  \bigl\langle 0\bigl| T\bigl[J^\mathrm{em}_\mu({\textstyle\frac{x}{2}})
  J^\mathrm{em}_\nu(-{\textstyle\frac{x}{2}})\bigr]\bigr|\pi\rangle
  \nn\\
  &&\hspace{0.5cm}\times \int_{-\infty}^{\infty} dt_y\int d^3\mathbf{y}\,\langle \pi|T[S^3(y) P^{(0)}(0)]|0\rangle.
  \nn\\
\ea
Here, $P^{(0)}$ does not have a quark-mass insertion. The associated quark-mass dependence has already been accounted for explicitly through the overall factor $\delta m^2$.

The integrated two-point function appearing in the expression for
$\Delta K^{\mathrm{IB},(2)}$ above is governed by the isoscalar--pion
overlap $\epsilon_\pi\equiv
\langle0|P^0|\pi\rangle_{\rm IB}/\langle0|P^3|\pi\rangle_{\rm IS}$, for
which
\be
\int d^4y\,\langle\pi|T[S^3(y)P^0(0)]|0\rangle
=-\epsilon_\pi\,\frac{\langle\pi|P^3(0)|0\rangle}{\delta m}.
\ee
Comparing with the pion-pole (first) term of Eq.~(\ref{eq:IS}),
$\Delta K^{\mathrm{IS}}_1$, then gives
\be
\Delta K^{\mathrm{IB},(2)}
=-\frac{\delta m}{2m_l}\,\epsilon_\pi\,\Delta K^{\mathrm{IS}}_1.
\ee
Using the direct 24D estimate $\epsilon_\pi=0.0067(25)$, together with
$\Delta K^{\mathrm{IS}}_1/K_0=-1.580(58)\%$, this gives
$\Delta K^{\mathrm{IB},(2)}/K_0=-0.0039(15)\%$.  In $SU(2)$ chiral
perturbation theory the overlap ratio is also related to the low-energy
constant $\ell_7$~\cite{Frezzotti:2021ahg,Bonanno:2025cmh}.

\subsection{Intermediate-state reconstruction}
\label{sec:intermediate}

The long-distance part of the leading isospin-breaking (IB) contribution
$\Delta K^{\mathrm{IB},(0)}$ is reconstructed from the $\eta$ and $\eta'$
intermediate states. 

\begin{figure}[h]
\centering
\includegraphics[width=\columnwidth]{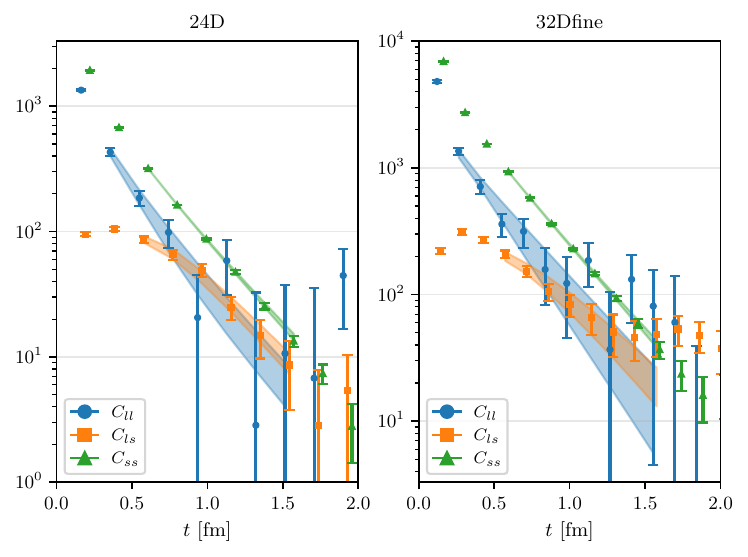}
\caption{The two-point correlation functions $C_{ll}(t)$, $C_{ls}(t)$, and $C_{ss}(t)$ constructed from the Coulomb-gauge-fixed wall-source operators $O_l$ and $O_s$, together with the results of the combined two-state fit, yielding the masses $m_\eta$ and $m_{\eta'}$ and the matrix elements $Z_{l,s}^{\eta^{(\prime)}}=\langle0|O_{l,s}|\eta^{(\prime)}\rangle$.}
\label{fig:eta_fit}
\end{figure}

We first determine the $\eta$ and $\eta'$ masses using 
Coulomb-gauge-fixed wall-source operators. We tested pseudoscalar and 
axial-current operators for both the light- and strange-quark sectors and found 
that the light-quark axial-current and strange-quark pseudoscalar operators provide the best signals. We therefore define
$O_l=\frac{1}{\sqrt2}(\bar u\gamma_4\gamma_5 u+\bar d\gamma_4\gamma_5 d)$ and $O_s=\bar s\gamma_5 s$.
A combined fit to the corresponding two-point correlation functions, shown in Fig.~\ref{fig:eta_fit}, determines the $\eta$ and $\eta'$ masses. 
Using the resulting matrix elements $Z_{l,s}^{\eta^{(\prime)}}=\langle0|O_{l,s}|\eta^{(\prime)}\rangle$, we construct optimized interpolating operators
\be
O_{\eta}^{\mathrm{opt}}=Z_s^{\eta'}O_l-Z_l^{\eta'}O_s,\quad O_{\eta'}^{\mathrm{opt}}=Z_s^{\eta}O_l-Z_l^{\eta}O_s.
\ee
The fit window used to determine the $\eta$ and $\eta'$ masses and matrix elements is indicated in Fig.~\ref{fig:eta_fit}, with the extracted masses given in Table~\ref{tab:eta_etap_mass_free}.

\begin{table}[h]
\centering
\begin{tabular}{lccc}
\hline
 & $m_\eta$ [GeV] & $m_{\eta'}$ [GeV] & $\chi^2/\mathrm{dof}$ \\
\hline
24D      & 0.585(26) & 1.11(15) & 0.387 \\
32Dfine  & 0.578(51) & 0.87(12) & 0.724 \\
PDG & 0.54786(2) & 0.95778(6) & -- \\
\hline
\end{tabular}
\caption{Masses of the $\eta$ and $\eta'$ mesons extracted from a two-state fit to the two-point correlation functions.}
\label{tab:eta_etap_mass_free}
\end{table}

\begin{figure}[h]
\centering
\includegraphics[width=\columnwidth]{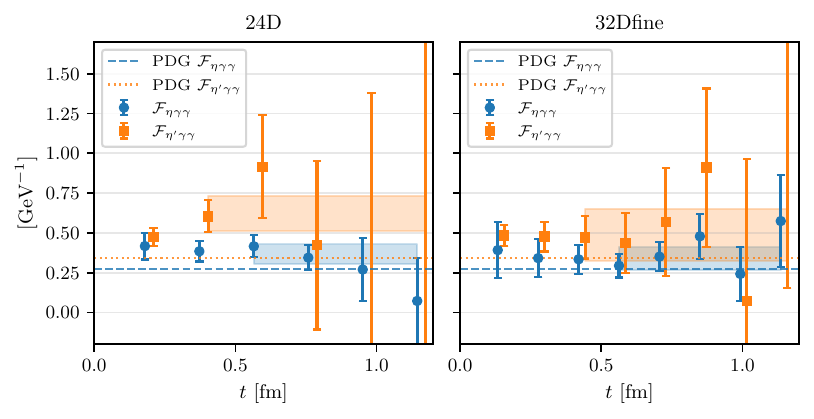}
\caption{Transition form factors for $\eta\to\gamma\gamma$ and $\eta'\to\gamma\gamma$ as a function of $t$, with $t=\min\{t_{J_1},t_{J_2}\}-t_{\mathrm{opt}}$ denoting the temporal separation between the pseudoscalar operator and the two electromagnetic-current insertions.}
\label{fig:eta_tff}
\end{figure}

As a next step, we use the optimized operators to determine the $\eta\to\gamma\gamma$ and $\eta'\to\gamma\gamma$ transition form factors. 
Fig.~\ref{fig:eta_tff} shows the time dependence of the corresponding transition form factors, with $t=\operatorname{min}\{t_{J_1},t_{J_2}\}-t_{\mathrm{opt}}$,
where $t_{J_1}$ and $t_{J_2}$ denote the insertion times of the two electromagnetic currents. We enlarge $t$ to suppress excited-state contamination. 
The $\eta'\to\gamma\gamma$ transition form factor has a larger statistical uncertainty, but its contribution to $\Delta K/K_0$ is only $\sim 6\%$ and $\sim 8\%$ 
on the 24D and 32Dfine ensembles, respectively, making its impact on the final result marginal.
The resulting transition form factors are obtained from constant fits and are summarized in Table~\ref{tab:TFF}.

\begin{table}[h]
\centering
\begin{tabular}{lccc}
\hline
 & $F_{\eta\gamma\gamma}$ [GeV$^{-1}$] & $F_{\eta'\gamma\gamma}$ [GeV$^{-1}$] & $\chi^2/\mathrm{dof}$ ($\eta/\eta'$) \\
\hline
24D      & 0.367(61) & 0.623(112) & 0.67/0.40 \\
32Dfine  & 0.339(71) & 0.488(163) & 0.59/0.35 \\
PDG & 0.2746(18) & 0.3413(76) & -- \\
\hline
\end{tabular}
\caption{Transition form factors  of $\eta\to\gamma\gamma$ and $\eta'\to\gamma\gamma$ extracted from the three-point correlation functions.}
\label{tab:TFF}
\end{table}

\begin{figure}[h]
\centering
\includegraphics[width=\columnwidth]{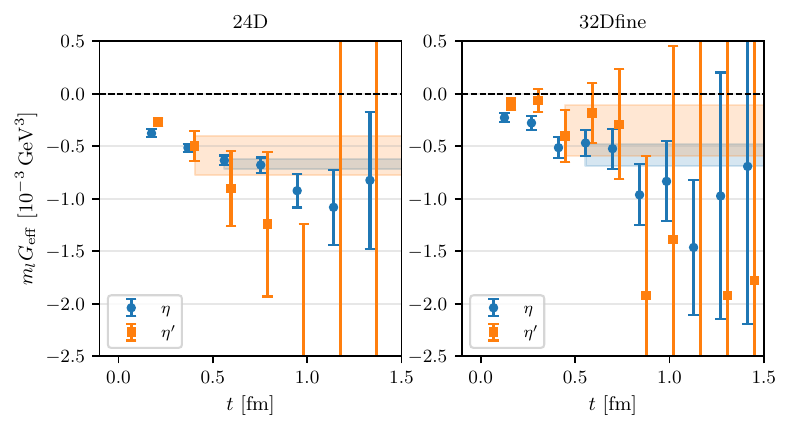}
\caption{Matrix elements $G_{\eta^{(\prime)}}=\langle \eta^{(\prime)}|P^0|0\rangle$ from two-point correlation functions constructed 
from $P^0$ and the optimized operators $O_{\eta^{(\prime)}}^{\mathrm{opt}}$.}
\label{fig:G_eta}
\end{figure}

Finally, we determine the matrix elements $G_{\eta^{(\prime)}}=\langle \eta^{(\prime)}|P^0|0\rangle$ from two-point correlation functions constructed 
from $P^0$ and the optimized operators $O_{\eta^{(\prime)}}^{\mathrm{opt}}$. The corresponding time dependence is shown in Fig.~\ref{fig:G_eta}. We multiply the matrix elements by the light-quark mass to eliminate the dependence on the renormalization factor of the pseudoscalar density. The final results are obtained from fits over the plateau region and are summarized in Table~\ref{tab:G}.

\begin{table}[h]
\centering
\begin{tabular}{lccc}
\hline
 & $m_l G_{\eta}$  & $m_l G_{\eta'}$ & $\chi^2/\mathrm{dof}$ ($\eta/\eta'$) \\
\hline
24D      & $-0.670(50)$ & $-0.59(19)$ & 0.97/0.73 \\
32Dfine  & $-0.58(10)$ & $-0.35(24)$ & 0.80/0.39 \\
\hline
\end{tabular}
\caption{Matrix elements $G_{\eta^{(\prime)}}=\langle \eta^{(\prime)}|P^0|0\rangle$ extracted from the two-point correlation functions. The matrix elements are given in units of $10^{-3}$ GeV$^{3}$.}
\label{tab:G}
\end{table}

\subsection{Finite-volume effects}

\begin{figure}[h]
\centering
\includegraphics[width=\columnwidth]{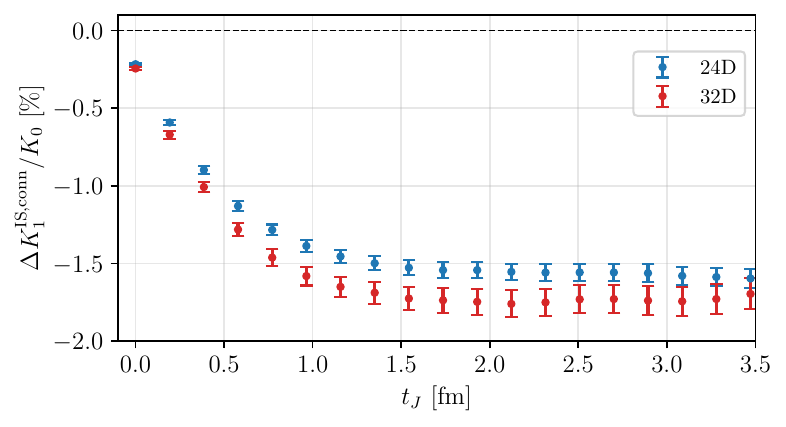}
\caption{Lattice results of $\Delta K_1^{\mathrm{IS,conn}}/K_0$
for 24D and 32D.}
\label{fig:24D_vs_32D}
\end{figure}

An earlier RBC-UKQCD study~\cite{Christ:2022rho} investigated the FV dependence of $\pi^0\to e^+e^-$ and $\pi^0\to\gamma\gamma$ by comparing the 24D and 32D ensembles, which have otherwise identical parameters but spatial extents corresponding to $L=4.2$ and $5.6$ fm, respectively. 
A FV effect of about $3\%$ was found for $\pi^0\to\gamma\gamma$ amplitude. The dominant contribution to our result comes from $\Delta K^{\rm IB}$, 
whose long-distance part is dominated by $\eta$ and $\eta'$ exchange and is therefore expected to be less sensitive to FV effects. Even conservatively assuming a $3\%$ FV effect, 
its impact is well below the current $10$-$15\%$ uncertainties on the individual ensembles and the $\sim50\%$ uncertainty after the continuum extrapolation. 
For $\Delta K^{\rm IS}$, the long-distance contribution is instead dominated by pion exchange. Moreover, while the IB contribution depends only on $\hat{\mathcal K}^{(0)}(0)$, the IS contribution involves the difference $\hat{\mathcal K}^{(3)}(m_\pi^2)-\hat{\mathcal K}^{(3)}(0)$, potentially enhancing the relative FV effect through the subtraction. We therefore explicitly evaluated $\Delta K_1^{\rm IS}$ on the 32D ensemble and found a FV effect at the $10\%$ level (See Fig.~\ref{fig:24D_vs_32D}), corresponding to only about a $6\%$ relative effect on $\Delta K/K_0$. This remains subleading at our current level of precision.

\subsection{Distinction between pion decay constants}
A further source of uncertainty is the decay constant entering Eq.~(\ref{eq:decay_width_pion_pole}).
The decay width depends on $F_{\pi^0}$, whereas the experimentally determined $F_{\pi^\pm}$ is conventionally used.
The decay constant $F_{\pi}$ used in this work is $F_{\pi^0}$, with leading order effects from QED interaction and up/down-quark mass difference, is defined as
\ba
\langle 0 | A^3_\mu(0) | \pi^0(\vec p) \rangle = F_{\pi} p_\mu,
\ea
which implies
\ba
\langle 0 | \hat P^3(0) + \hat P^0(0) | \pi^0(\vec p) \rangle = -i F_{\pi} m_\pi^2,
\ea
where the operators $A^3_\mu, \hat P^3, \hat P^0$ are defined near \cref{eq:pcac}. We repeat them here
\begin{equation}
\begin{aligned}
A_\mu^3 =& \frac{1}{2}(\bar u\gamma_\mu\gamma_5 u - \bar d\gamma_\mu\gamma_5 d),
\\
\hat{P}^3=&2m_lP^3,
\\
\hat{P}^0=&\delta m P^0,
\end{aligned}
\end{equation}
where
\begin{equation}
\begin{aligned}
P^3 =& \frac{1}{2}(\bar u \gamma_5 u - \bar d \gamma_5 d),
\\
P^0 =& \frac{1}{2}(\bar u \gamma_5 u + \bar d \gamma_5 d),
\end{aligned}
\end{equation}
and
\begin{equation}
\begin{aligned}
m_l=&(m_u{+}m_d)/2,
\\
\delta m=&m_u-m_d.
\end{aligned}
\end{equation}
The leading QCD IB correction is suppressed to $\mathcal{O}(\delta m^2/\Lambda_\mathrm{QCD}^2)$ by the $m_u\leftrightarrow m_d$ symmetry
of both decay constants, yielding $F_{\pi^0}/F_{\pi^\pm}-1 = 2.1\times 10^{-4}$ in two-loop ChPT~\cite{Amoros:2001cp}.
Electromagnetic effects arise at $\mathcal{O}(e^2p^2)$; resonance-saturation estimates give
$|\delta_{\rm EM}F_\pi|\sim0.1\,\mathrm{MeV}$, corresponding to a relative effect of order $10^{-3}$~\cite{Neufeld:1995mu,Baur:1996ya}.
These estimates are therefore too small to account for the tension with experiment, but they rest on ChPT convergence and model input; a direct lattice determination of $F_{\pi^0}$, including its electromagnetic corrections, would be needed to remove this model dependence at the sub-percent level.

\end{document}